\documentstyle{article}
\begin{document}
\title{Umklapp Scattering and Heat Conductivity of Superlattices}
\author{M.V. Simkin and G.D. Mahan\\Department of Physics and Astronomy, University of Tennessee,\\
Knoxville, 37996-1200, and\\
Solid State Division, Oak Ridge National Laboratory,\\ P.O. Box 2008, 
Oak Ridge, TN, 37831}
\date{}
\maketitle
\begin{abstract}

The mean free path of phonons in superlattices is estimated.
It is shown to be strongly dependent on the superlattice period due to the
Umklapp scattering in subbands. It first falls with increasing the 
superlattice period until it becomes comparable with the latter after
what it rises back to the bulk value. Similar behavior is expected
of heat conductivity, which is proportional to the mean free path.
\end{abstract}
Superlattices offer an opportunity to control physical properties in 
unprecedented ways. Their thermal conductivity is of interest both for a 
fundamental understanding of these  systems as well as in applications.  
Recently there has been a resurgence of interest in finding materials with 
improved thermoelectric transport properties for cooling and power generation.
The quality of a material for such applications is given by the thermoelectric
figure of merit, which is inversely proportional to the thermal conductivity
$\kappa$.  In materials of interest, such as semiconductors, the lattice 
contribution to $\kappa$ dominates.

Experimental and theoretical work suggests that the thermal conductivity of 
superlattices is quite low,
both for transport along the planes \cite{yao,hyl2,rein}, or perpendicular 
to the planes \cite{mar1,lee,ven,chen,hyl1,mar2,sim}. 

The lattice heat conductivity $\kappa$ is given approximately by an equation 
\cite{zim}: 
\begin{eqnarray}
\kappa \approx Cvl,
\label{heat}
\end{eqnarray}
where $C$  is the lattice heat capacitance, $v$ - the average phonon group
velocity, and $l$ - the mean free path.
Recently we presented calculations of the thermal conductivity perpendicular 
to the layers \cite{sim} which were done in approximation which takes into 
account changes in phonon group velocities due to band folding, but neglects
the dependence of the phonon mean free path on the superlattice period.
The investigation of this dependence is the subject of the present work.

Three-phonon scattering due to anharmonicity is the dominant contribution to 
the lattice thermal resistivity.  Umklapp processes, in which the net phonon 
momentum change by a reciprocal lattice vector, give the finite thermal 
conductivity \cite{zim}.   Only phonons with energies of the order of Debye 
energy, $\Theta_D$ can participate in Umklapp scattering, giving a 
temperature dependence of the phonon mean free path $l$ of the form \cite{zim} 
\begin{eqnarray}
l \approx \exp(\Theta_D/T).
\label{l}
\end{eqnarray}
In superlattices, new mini-bands are introduced in the acoustic phonon 
dispersions along the growth direction, and they give rise to new Umklapp 
processes.  The lowest phonon energy for Umklapp scattering in a superlattice 
of period L is of order $\Theta_D/L$ and phonon mean free path in a superlattice, $L_{sl}$  becomes:
\begin{eqnarray}
l_{sl} \approx \exp(-\Theta_D/LT).
\label{lsl}
\end{eqnarray}
The ratio of the mean free paths is:
\begin{eqnarray}
l_{sl}/l \approx \exp(\Theta_D/T(1/L-1)),
\label{ratio}
\end{eqnarray}
which can be rather small for big $L$ and small $T$. 

According to Eq.\ref{ratio} $l_{sl}$ decreases with $L$ and eventually
should become $l_{sl}=L$. This shall happen at the value of $L=L_c$  given by a
solution to the equation:
\begin{eqnarray}
L_c/l = \exp(\Theta_D/T(1/L_c-1))
\label{Lc}
\end{eqnarray}
What happens next?  When $L$ exceeds $L_c$ according to 
Eq.\ref{ratio} it should become $l_{sl}<L$.
However in this case Eq.\ref{ratio} is no longer
applicable because when  $l_{sl}<L$ superlattice effects should not matter
and $l_{sl}$ should assume the bulk value $l$. But as $L<l$ then Eq.\ref{ratio}
should be valid again. The only resolution of this contradiction is that
$l_{sl}$ starts to increase with $L$  as $l_{sl} \approx L$ after reaching a 
minimum at $L=L_c$. It shall saturate, however when $L>l$
(in this case we are not bound to use Eq. \ref{ratio} again ($L<l$!).

Eq.\ref{Lc} is not soluble analytically but assymptotics are easy to 
compute. For large $l$ we get:
\begin{eqnarray}
L_c/l = \exp(-\Theta_D/T),
\label{La}
\end{eqnarray}
which can be orders of magnitude small.
When $l=1$ we get $L_c=1$.

The heat conductivity is proportional to $l_{sl}$  (Eq.\ref{heat}) and  
should follow its behavior.


\begin{thebibliography}{99}
\bibitem{yao} T. Yao, Appl. Phys. Lett. {\bf 51}, 1798 (1987)
\bibitem{hyl2} P. Hyldgaard and G.D. Mahan, Proc.Int. Conf. on Thermal 
Conductivity, Nashville,(November, 1995)
\bibitem{mar1} W.S. Capinski and H.J. Maris, Physica B {\bf 219-220}, 699 
(1996)
\bibitem{lee} S.M. Lee, D.G. Cahill, R. Ventakasubramanian, Appl. Phys. Lett. 
{\bf 70},
2957 (1997)
\bibitem{ven} R. Venkatasubramanian and T. Colpitts, in {\it Thermoelectric Materials--New
Directions and Approaches}, ed. T.M. Tritt, M.G. Kanatzidis, H.B. Lyons Jr, and G.D. Mahan
(Materials Research Society, 1997) Vol. 478, pg. 73;
R. Ventakasubramanian, Phys. Rev. B {\bf 61}, 3091 (2000)
\bibitem{chen} G. Chen and M. Neagu, Appl. Phys. Lett. {\bf 71}, 2761 (1997)
\bibitem{hyl1} P. Hyldgaard and G.D. Mahan, Phys. Rev. B {\bf 36}, 10754
(1997)
\bibitem{mar2} S. Tamura, Y. Tanaka, and H.J. Maris, Phys. Rev. B  {\bf 60},
2627 (1999)
\bibitem{chen2} G. Chen, Phys. Rev. B {\bf 57}, 14958 (1998)
\bibitem{rein}S. G. Walkauskas, D. A. Broido, K. Kempa, and T. L. Reinecke, 
J. Appl. Phys {\bf 85}, 2579 (1999)
\bibitem{sim} M. V. Simkin and G. D. Mahan, Phys. Rev. Lett. 84, 927 (2000).
\bibitem{zim} J. Ziman {\it Electrons and Phonons} (Cambridge University Press, 1960) 
\end{thebibliography}
\end{document}